\documentclass[aps,prl,preprint,showpacs,superscriptaddress]{revtex4}
\usepackage{amsmath}
\usepackage{amsfonts}
\usepackage{amssymb}
\usepackage{graphicx,epsfig}

\newcommand{\ZZ}{\mathcal{Z}}
\newcommand{\FF}{\mathcal{F}}
\newcommand{\ud}{\mathrm{d}}
\newcommand{\kb}{k_{\mathrm{B}}}
\newcommand{\mub}{\mu_{\mathrm{B}}}
\newcommand{\EF}{E_{\mathrm{F}}}
\newcommand{\Selec}{\bar{S}_{elec}}
\newcommand{\Smag}{\bar{S}_{mag}}
\newcommand{\Slatt}{\bar{S}_{latt}}

\begin{document}
%\draft
\title{Fluctuating local moments, itinerant electrons and the magnetocaloric effect: the compositional hypersensitivity of FeRh}
\author{J. B. Staunton, R. Banerjee}
\affiliation{Department of Physics, University of Warwick, Coventry CV4 7AL, U.K.}
\author{M. dos Santos Dias}
\affiliation {Peter Gr\"unberg Institut and Institute for Advanced Simulation, Forschungszentrum J\"ulich and JARA,
D-52425 J\"ulich, Germany}
\author{A. Deak}
\affiliation{Department of Theoretical Physics, Budapest University of Technology and Economics, Budapest, Hungary}
\author{L. Szunyogh}
\affiliation{Condensed Matter Research Group of the Hungarian Academy of Sciences,  
Budapest University of Technology and Economics, Budapest, Hungary}
\date{\today}

\begin{abstract}
We describe an ab-initio Disordered Local Moment Theory for materials with quenched static compositional disorder traversing first order magnetic phase transitions.
It accounts quantitatively for metamagnetic changes and the magnetocaloric effect.
For perfect stoichiometric B2-ordered FeRh, we calculate the transition temperature of the ferromagnetic-antiferromagnetic 
transition to be $T_t =$ 495K and a maximum isothermal entropy change in 2 Tesla of  $|\Delta S|= 21.1$ J~K$^{-1}$~kg$^{-1}$.
A large (40\%) component of $|\Delta S|$ is electronic.
The transition results from a fine balance of competing electronic effects which is disturbed by small compositional changes - 
e.g. swapping just 2\% Fe of `defects' onto the Rh sublattice makes $T_t$ drop by 290K.
This hypersensitivity explains the narrow compositional range of the transition and impurity doping effects.
\end{abstract}
%\pacs{75.30.Kz, 75.10.Lp,75.50.Ee,75.30.Sg,75.80.+q}
\maketitle

When a metal goes through a change of magnetic order, the complex electronic fluid with its emergent magnetic fluctuations transforms.
The magnetic effect on structure, electronic transport and so on is particularly dramatic at first order phase transitions, and modest changes in composition and quenched disorder are strongly influential, shifting, broadening~\cite{Imry+Wortis,Roy2008} or removing transitions entirely.
In a metallic material the changes to the electronic structure as it passes through a first order magnetic transition can be significant. In this paper we show how these changes can be determined, find out how they are affected by composition and disorder and trace their impact back on the transition itself.
We find a particularly striking example in the metamagnetic Fe-Rh material.e.g.~\cite{Fallot,Annaorazov1996,vanDriel1999,Kobayashi2001,Maat2005,Stamm2005,Suzuki2011,Marrows2013}. 

In roughly equal proportions, iron and rhodium order into a B2 (CsCl) alloy phase which experiences a first order ferromagnetic (F) to antiferromagnetic (AF) phase transition at $T_t$ around 340K, below a $T_c$ of 670K.
This property is highly composition-dependent so that the F-AF transition vanishes in alloys with as little as a 2\% iron excess or deficiency and 
$T_t$ varies strongly with sample preparation, irradiation and addition of impurities~\cite{Tu1969,Hashi2004,Cao2008,Kosugi2011,Kouvel1966,Cooke2012,Lewis2013}.
For example a little Pd raises $T_t$ whilst doping  with Pt suppresses it.
The 1\% Fe-deficient alloy shows one of the largest recorded magnetocaloric effects around $T_t$~\cite{Annaorazov1992a,Annaorazov1996} which deteriorates on subsequent magnetic and thermal cycling.
The prominent F-AF transition is also relevant for the design of ultra high density magnetic recording media as shown by 
FePt/FeRh bilayer investigations\cite{Thiele2003} and its time-dependence has also been probed by a recent suite of experiments
and analysis~\cite{Ju2004,Sandratskii2011,Mariager2012,Derlet}.

The electronic source of the F-AF transition has been tracked down carefully by several spin density functional theory (SDFT)-based 
studies~\cite{Moruzzi1992,Gruner2003,Sandratskii2011,Mryasov2005,Derlet}, which have also looked at the role
of spin-waves~\cite{Gu2005}, and much insight gained.
For an AF state with zero sum magnetisation on the Fe sites, the Rh related states have no net spin-polarisation and there is a rough half-filling of some of the Fe-related $d$-bands, which sustains the long-range AF order~\cite{Kubler-book}.
Around the Fermi energy, $\EF$, there is a strong hybridisation between Fe and Rh states in both spin channels for both F or AF order.
In the former case, however, where the Rh sites pick up an overall
spin-polarisation~\cite{Mryasov2005,Gruner2003}, some bonding states are pulled down in energy enhancing the F state.
The balance between these two competing effects drives the F-AF transition.
We show in this letter, by allowing for finite $T$ magnetic fluctuation effects, that tiny alterations of composition and  quenched disorder change the electronic structure to tip this balance which makes $T_t$ and the critical fields $H_c$ triggering the transition highly composition-sensitive. 

Long-range compositional order is never perfect in any real Fe-Rh sample and there can also be a slight off-stoichiometry.
Given that the alloy orders from a Fe$_{50}$Rh$_{50}$ solid solution around 1600K~\cite{vanDriel1999}, a simple Bragg-Williams model~\cite{BW} analysis indicates that at 
least 1 or 2\% of the sites on the Rh cubic sub-lattice will be occupied by Fe and vice-versa following typical annealing and cooling processes.
The number of Fe-occupied Rh sub-lattice sites diminishes for marginally Fe-poor alloys.
Our theoretical work finds the F-AF transition to be profoundly influenced by such Fe anti-site defects and explains the narrow range of composition for the transition.
Added impurities affect the transition by how easily they displace Fe atoms onto such defects. 

Important magnetic fluctuations in a metal can often be modelled as `local moments', a picture  captured by a generalisation of SDFT~\cite{blg} for non-collinear spin-polarisation.
A separation of timescales between fast and slow electronic degrees of freedom causes `local moments' with slowly varying orientations, $\{\hat{e}_i\}$ to emerge from the interacting electron system ~\cite{jh,hh,JBS+BLG,RDLM-FePt,Manuel,Nature,TM-oxides}.
This means that `disordered local moments' (DLM)  are sustained by and influence the faster electronic motions. Their interactions with  each other depend on the type and extent of the long range magnetic order through the associated spin-polarized 
electronic structure~\cite{CoMnSi} which itself adapts to the extent of magnetic order.
Ensemble averages over all the appropriately weighted non-collinear local moment orientational configurations, $\{\hat{e}_i\}$, are required for a realistic evaluation of the system's magnetic properties~\cite{RDLM-FePt,RDLM-big}.
Here we develop the DLM theory for a magnetic material in an external magnetic field $\vec{H}$ at a temperature $T$ for application to systems with quenched disorder like our FeRh case study.
For the first time  we show how entropy changes that occur at magnetic transitions, the magnetocaloric effect (MCE), can be calculated ab-initio, quantify the potentially significant electronic 
contribution and its signature in temperature dependent magnetotransport properties~\cite{Marrows2013}.

For an Fe-Rh alloy close to equiatomic stoichiometry and nearly complete B2-type order there are two atomic sites per unit cell in the cubic crystal lattice.
One sub-lattice (A) has sites, labelled $a$,  largely Fe-occupied but with a small percentage ($x$) of sites occupied by Rh atoms.
The other sublattice (B), with sites $b$, mostly occupied by Rh atoms has a small fraction ($y$) Fe-occupied. The alloy is designated Fe$_{1-x}$Rh$_x$--Rh$_{1-y}$Fe$_y$.
A particular distribution of Fe and Rh atoms over the two sublattices is specified with $(\{\xi_a\}$, $\{\xi_b\})$, where $\xi_{a(b)}=\{0,1\}$ means site $a(b)$ is \{Rh,Fe\} occupied, such that $\langle\xi_{a}\rangle = c_A = 1 - x$, and likewise $\langle\xi_{b}\rangle = c_B = y$. 
The probability that the system's local moments are configured according to $\{\hat{e}_a\}$, $\{\hat{e}_b\}$ is $P(\{\hat{e}_a\},\{\hat{e}_b\}) = \exp[-\beta\Omega(\{\hat{e}_a\},\{\hat{e}_b\})]/\ZZ$, where the partition function is $\ZZ = \prod_a\!\int\!\ud\hat{e}_a \prod_b\!\int\!\ud\hat{e}_b\,\exp[-\beta\Omega(\{\hat{e}_a\},\{\hat{e}_b\})]$, $1/\beta = \kb T$ and the free energy $\FF = -\kb T \ln\ZZ$.
A `generalised' electronic grand potential $\Omega(\{\hat{e}_a\},\{\hat{e}_b\};\vec{H},\{\xi_a\},\{\xi_b\},T)$ is in principle available from SDFT~\cite{blg} where, for fixed $\vec{H}$ and for the arrangement of atoms specified by $\{\xi_a\}$ and $\{\xi_b\}$, the spin density is constrained to comply with the local moment configurations $\{\hat{e}_a\}$ and $\{\hat{e}_b\}$. 
It thus plays the role of a local moment Hamiltonian but its electronic glue origins can make it complicated. 

Expanding about a suitable reference `spin' Hamiltonian $\Omega_{0} = \sum_{a} \vec{h}_a \cdot \hat{e}_a +\sum_{b} \vec{h}_b \cdot \hat{e}_b$~\cite{Feynman}, gives a mean field theoretical estimate of the free energy~\cite{blg}.
A similar single site approximation averages over atomic configurations with the assumption that atomic diffusion times are very long and that the composition is fixed by the material's preparation.
Local moments establish on the Fe atoms only and, in line with $T = 0$K DFT studies~\cite{Sandratskii2011} and consistent
with other theoretical studies~\cite{Gruner2003,Mryasov2005}, a net spin-polarisation develops on the Rh atoms if there is an overall lining up of the Fe local moments, i.e. when F order is established.
The free energy is given by
\begin{align}
\label{Free}
  \FF =  \bar{\Omega}
&+ c_A \sum_a \Big(\mu_{a}\,\vec{m}_{a}\cdot\vec{H} - \frac{1}{\beta}\!\int\!\ud\hat{e}\,P_{a}(\hat{e}) \ln P_{a}(\hat{e}) \Big) \nonumber \\
&+ c_B \sum_b \Big(\mu_{b}\,\vec{m}_{b}\cdot\vec{H} - \frac{1}{\beta}\!\int\!\ud\hat{e}\,P_{b}(\hat{e}) \ln P_{b}(\hat{e}) \Big)
\end{align}
The free energy therefore comprises an internal energy $\bar{\Omega}$ from the interacting electron system averaged over local moment and compositional configurations,
and two extra contributions from the Fe local moments in each sublattice: the interaction with the external magnetic field $\vec{H}$ and the magnetic entropy, $-T\,\bar S_{mag}$.
The magnitudes of the local moment on each site, $\mu_{a}$ and $\mu_{b}$, are determined by the generalised SDFT. 
$\bar{\Omega}$ includes the effect on the spin-polarized electron density at the Rh sites from $\vec{H}$ and the magnetic order of the Fe local moments~\cite{Lezaic2013,Deak}.
The probability of an Fe local moment being oriented along $\hat{e}_a$ on site $a$ of the Fe-rich sub-lattice is set as
$ P_{a}(\hat{e}_a) = \exp[\vec{\lambda}_{a}\cdot\hat{e}_a]\,/\!\int\!\ud\hat{e}_a \exp[\vec{\lambda}_{a}\cdot\hat{e}_a]$ and similarly for an Fe atom defect on the Rh-rich sub-lattice, $P_{b}(\hat{e}_b)$.
A magnetic state is specified by the set of local order parameters, $\{\vec{m}_{a} = \int\!\ud \hat{e}_a\,P_{a}(\hat{e}_a)\,\hat{e}_a
= \big(\!\coth(\lambda_{a}) - 1/\lambda_{a}\big)\hat{\lambda}_{a}\}$ and $\{\vec{m}_{b}\}$, each of which can take values between $0$ and $1$.
The parameters $\{\vec{\lambda}_{a}\}$, $\{\vec{\lambda}_{b}\}$ are given as $\{\beta\vec{h}_{a}\}$, $\{\beta\vec{h}_{b}\}$ respectively, where the Weiss fields satisfy $\vec{h}_{a}= -\frac{1}{c_A}\frac{\partial\bar\Omega}{\partial\vec{m}_{a}} - \mu_{a} \vec{H}$ and $\vec{h}_{b} = -\frac{1}{c_B}\frac{\partial\bar\Omega}{\partial\vec{m}_{b}} - \mu_{b} \vec{H}$.
This ensures that the function $\FF(\{\vec{m}_{a}\},\{\vec{m}_{b}\};\vec{H},c_A,c_B,T)$, shown in Eq.~\ref{Free}, is minimized with respect to the $\{\vec{\lambda}_{a}\}$, $\{\vec{\lambda}_{b}\}$ (equivalently $\{\vec{m}_{a}\}$, $\{\vec{m}_{b}\}$), at a temperature $T$, and hence describes the free energy.

Particularly pertinent for our description of MCE~\cite{sandeman_2012a} is the electronic entropy contained in $\bar\Omega$ of Eq. \ref{Free} ~\cite{Mermin}, $\bar{\Omega} = \bar{E} - T \, \Selec$.
$\bar E$ is the SDFT-based energy averaged over local moment orientations and compositional arrangements~\cite{RDLM-big,CoMnSi,KKRCPA,scf-kkr-cpa,KKR-review} with electronic density of states (DOS) at the Fermi energy $\bar{n}(\EF;\{\vec{m}_{a}\},\{\vec{m}_{b}\};\vec{H},c_A,c_B)$ and $\Selec$ is the electronic entropy.
$\Selec \approx \frac{\pi^2}{3}\,k_B^2 T\,\bar{n}(\EF)$ from the Sommerfeld expansion.
The isothermal entropy difference between states with and without a magnetic field applied, $\Delta S(\vec{H},T)$, is therefore comprised of the sum of the $\Smag$ and $\Selec$ differences.
Likewise the adiabatic temperature change, $\Delta T_{ad}(\vec{H},T)$ can be estimated from $\Smag(\vec{H},T) + \Selec(\vec{H},T) + \Slatt(T) = \Smag(\vec{0},T+\Delta T_{ad}) + \Selec(\vec{0},T+\Delta T_{ad}) + \Slatt(T+\Delta T_{ad})$ where $\Slatt$ is the lattice vibration entropy.

To examine the compositional sensitivity of the F-AF transition to disorder, we apply the theory to two magnetic states of a Fe$_{1-x}$Rh$_{x}$--Rh$_{1-y}$Fe$_y$ alloy at many temperatures, both with and without an external field $\vec{H} = H\hat{z}$. The first
state is a ferromagnetic state (F) with local moments on the Fe sites all set to $\vec{m}_{a} = m_{f}\hat{z}$, describing how aligned the Fe moments are on the Fe-rich sublattice (A), and all $\vec{m}_{b} = m_{f'}\hat{z}$, describing the analogous order parameter for the the anti-site Fe moments on the Rh-rich sublattice (B).
The second is a canted anti-ferromagnetic magnetic state (AF)~\cite{Derlet}, with order parameters $\vec{m}_{a}$ alternating between $m_{f}\hat{z} + m_{af}\hat{x}$ and $m_{f}\hat{z} - m_{af} \hat{x}$ on the two interleaved fcc sublattices which form the A sublattice, and all $\vec{m}_{b} = m_{f'}\hat{z}$ for the sites of the B sublattice.
The paramagnetic state is specified by $m_{f} = m_{f'} = m_{af} = 0$, and the $T = 0$K magnetic ground states by $m_{f} = m_{f'} = 1$, $m_{af} = 0$ for the F state and $m_{af}= 1$, $m_{f} = m_{f'} = 0$ for the AF state. In general $m_{f}$, $m_{f'}$ are ferromagnetic order parameters whilst $m_{af}$ describes the extent of 
anti-ferromagnetic order.

For specific concentrations $(x,y) \Leftrightarrow (c_A,c_B)$, and magnetic state (F or AF) we select many values of $m_{f}$, $m_{f'}$, $m_{af}$ (80-120 sets) and calculate ab-initio~\cite{RDLM-FePt,CoMnSi} 
$\bar{\Omega}$ and $\Selec$ averaged with the probability distributions consistent with these choices.
We find that in terms of $m_{f}$, $m_{f'}$ and $m_{af}$
\begin{align}
\label{fit}
  \bar{\Omega} \approx E_0 - & \frac{\pi^2}{6}(\kb T)^2 \bar{n}(\EF) \\- c_A\,\Big( e_{af} m_{af}^2  + & g_{af} m_{af}^4 + e_{f} m_{f}^2 + g_{f} m_{f}^4 
  + g_{aff} m_{af}^2 m_{f}^2\Big)\nonumber \\ - c_B\,\Big( e_{f'} m_{f'}^2 + & e_{ff'} m_{f} m_{f'} 
  + g_{aff'} m_{af}^2 m_{f'}^2 + g_{ff'} m_{af}^3 m_{f'}  \Big)\nonumber
\end{align}
and
\begin{equation}
  \label{elec}
  \bar{n}(\EF) \approx n_0 + c_A \big(n_{af} m_{af}^2 + n_{f} m_{f}^2\big) + c_B\,n_{ff'} m_{f} m_{f'}
\end{equation}
fit our ab-initio computational data very well. Eq.~\ref{elec} reflects how the spin-polarised electronic structure adapts 
to the extent and type of long range magnetic order.
All fit coefficients depend on $c_A$, $c_B$ and the magnetic state.
For the F state, $e_{af}$, $g_{af}$, $g_{aff}$, $g_{aff'}$, and $n_{af}$ coefficients are all zero and if $c_B = 0$, when there are no Fe atoms on sublattice B, those coefficients associated with the B sub-lattice are not needed.
The $e_{f}$ coefficient includes the effect derived from the spin polarisation on the Rh sites which is caused by the Fe local moments lining up ($m_{f}$)~\cite{Lezaic2013,Deak}.
We use Eqs.~\ref{fit} and \ref{elec} in Eq.~\ref{Free} to specify the free energy function, $\FF(m_{af},m_{f},m_{f'};c_A,c_B,\vec{H},T)$ and find, for both F and AF states, those values of $m_{af}$, $m_{f}$, and $m_{f'}$ which minimise it for selected $T$, $\vec{H}$, $c_A$ and $c_B$ values, i.e. $\bar\FF_{F}(\vec{H},c_A,c_B,T)$ and $\bar\FF_{AF}(\vec{H},c_A,c_B,T)$.
By comparing $\bar\FF_{F}$ and $\bar\FF_{AF}$ we can locate the F-AF transition, its magnetic field dependence and associated MCE and electronic structure changes.  

Our first application is to completely ordered Fe-Rh, ($x=y=0$, i.e. $c_A=1$,$c_B=0$) with the lattice spacing 3.0~\AA, in line with experiment.
A local moment of 3.15 $\mub$ forms on the Fe atoms and a net spin-polarisation is induced on Rh of 1.00 $\mub$ per atom, when the Fe moments are fully ferromagnetically aligned ($m_{f} = 1$) but vanishes when the Fe local moments are disordered, ($m_{f} = 0$).
The coefficients in Eqs.~\ref{fit} and \ref{elec} are calculated to be (in meV) $e_{af} = 87.6$, $e_{f} = 99.4$, $g_{af} = 18.2$, $g_{f} = -23.0$ and $g_{aff} = -35.6$.
$e_{f} > e_{af}$ so that in zero field the material orders ferromagnetically at $T_c = 773$K whilst the negative signs of $g_{f}$ and $g_{aff}$ ensure a transition to an AF state at $T_t = 495$K  (experimental values are $T_c = 670$K and $T_t = 340$K).
There is a large change to $\bar{n}(\EF)$ at $T_t$  (i.e. change of $(n_{af} m_{af}^2 + n_{f} m_{f}^2)$~\cite{Suzuki2011} with $n_{af} = -1.3$ and $n_{f} = 0.59$ states/eV/FeRh pair) where the order parameters are $m_{f} = 0.59$, $m_{af} = 0$ above $T_t$ and $m_{f} = 0$, $m_{af} = 0.71$ below.
We include the effects of an applied field $\vec{H}$ and find $\frac{dT_t}{dH} = -3.7$ K~T$^{-1}$. The isothermal entropy change for 2 Tesla has a maximum value at the transition of $|\Delta S| = 21.1$ J~K$^{-1}$~kg$^{-1}$ which compares well with values reported experimentally of 
$|\Delta S| = 13$ - $20$ J~K$^{-1}$~kg$^{-1}$~\cite{Annaorazov1996,Cooke2012,sandeman_2012a}.
Notably $|\Delta S|$ has a large electronic component --- reducing  to 13 J~K$^{-1}$~kg$^{-1}$ if $\Selec=\frac{\pi^2}{3}\,k_B^2 T\,\bar{n}(\EF)$ (see Eq.~\ref{fit}) is omitted.  

\begin{figure}[h]
\begin{center}
\includegraphics[width=42mm,height=36mm]{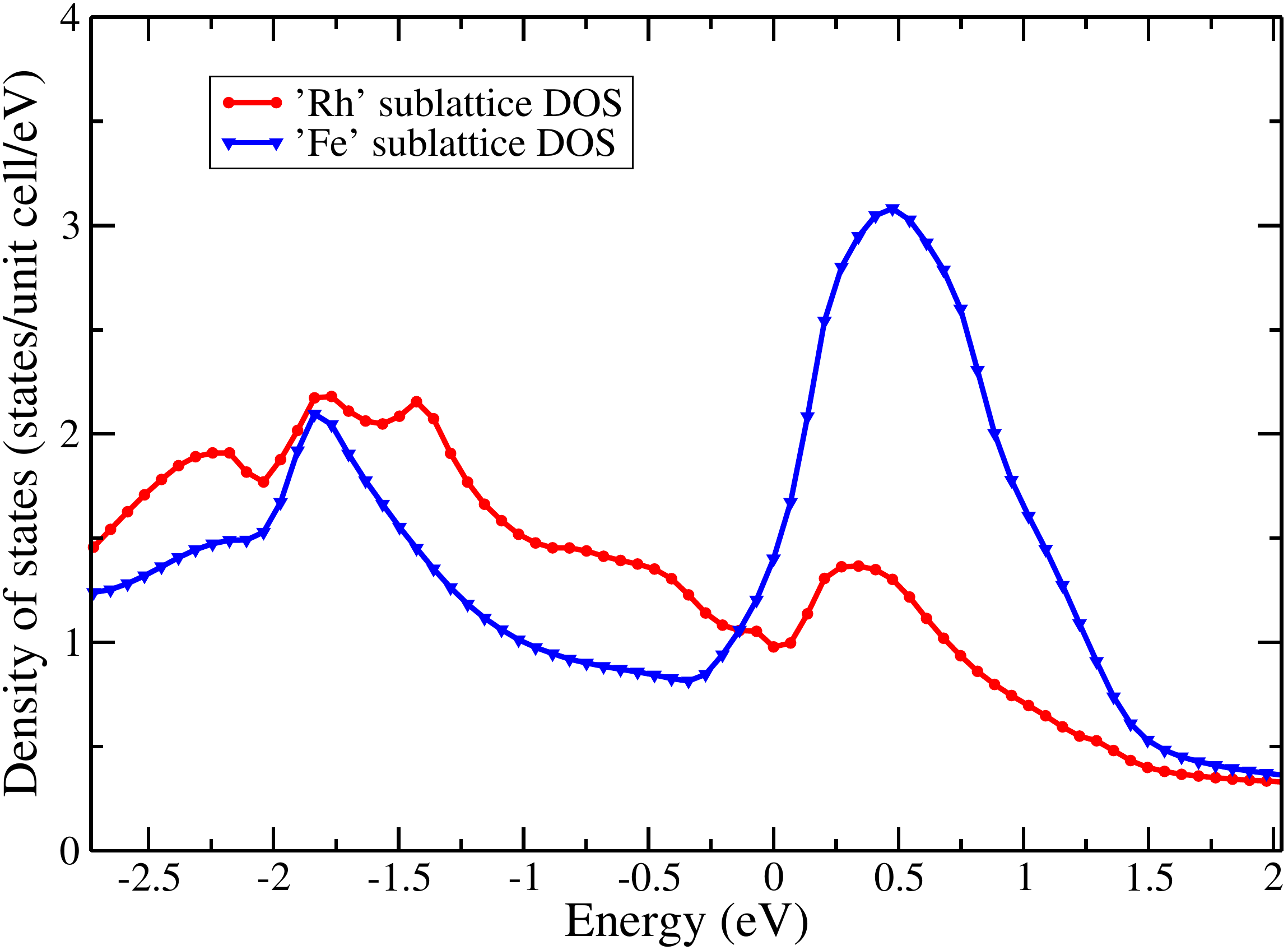}
\includegraphics[width=42mm,height=36mm]{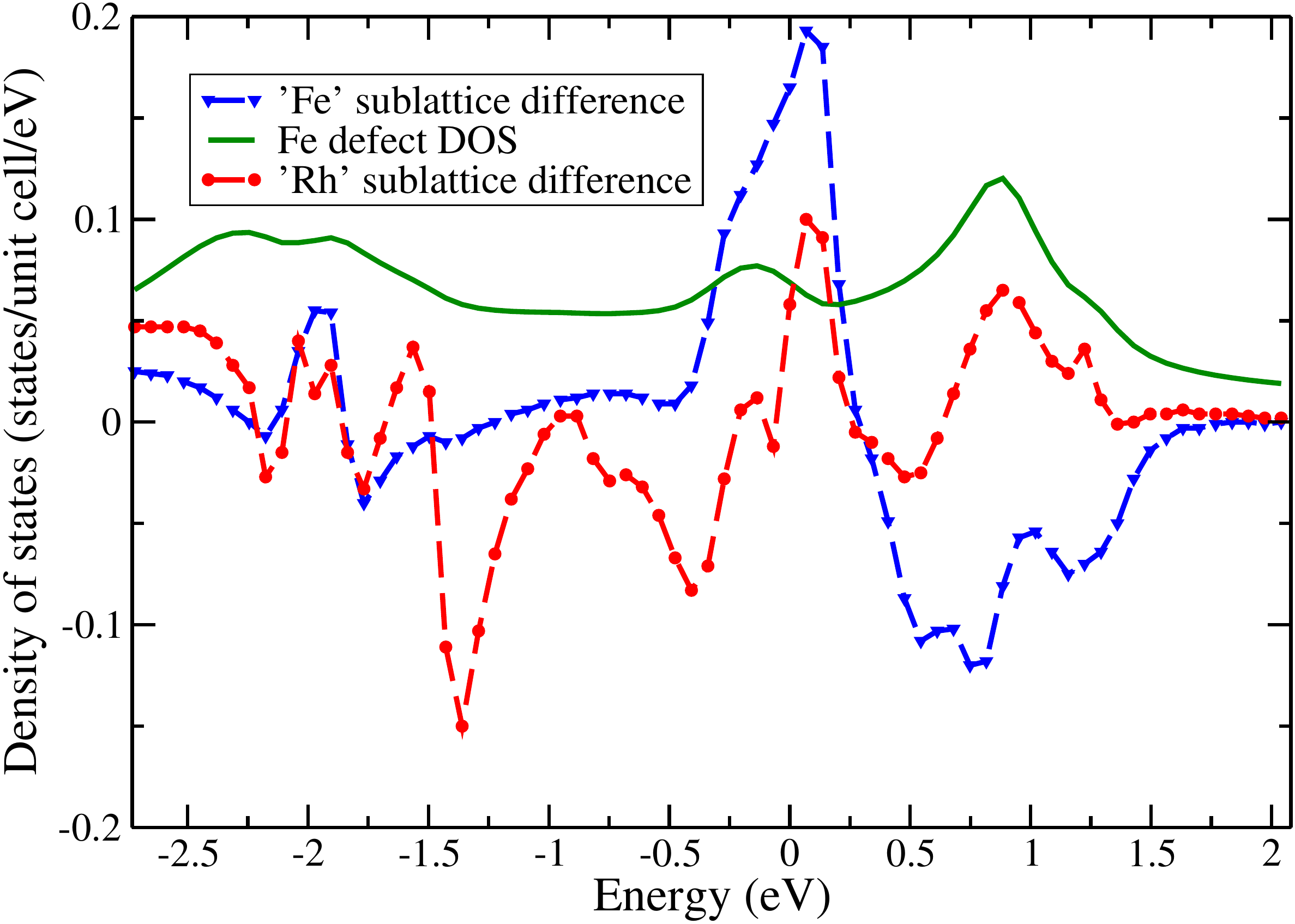}
\caption{(a) Density of states of Fe-Rh above $T_c$ resolved into sublattice components.
(b) The difference in the DOS between Fe-rich Fe-Rh$_{0.96}$Fe$_{0.04}$ and Fe-Rh above $T_c$. 
The Fe `defect' DOS on the Rh sublattice is also shown.}
\label{Fig2}
\end{center}
\end{figure}
\begin{figure}[h]
\begin{center}
\includegraphics[width=42mm,height=36mm]{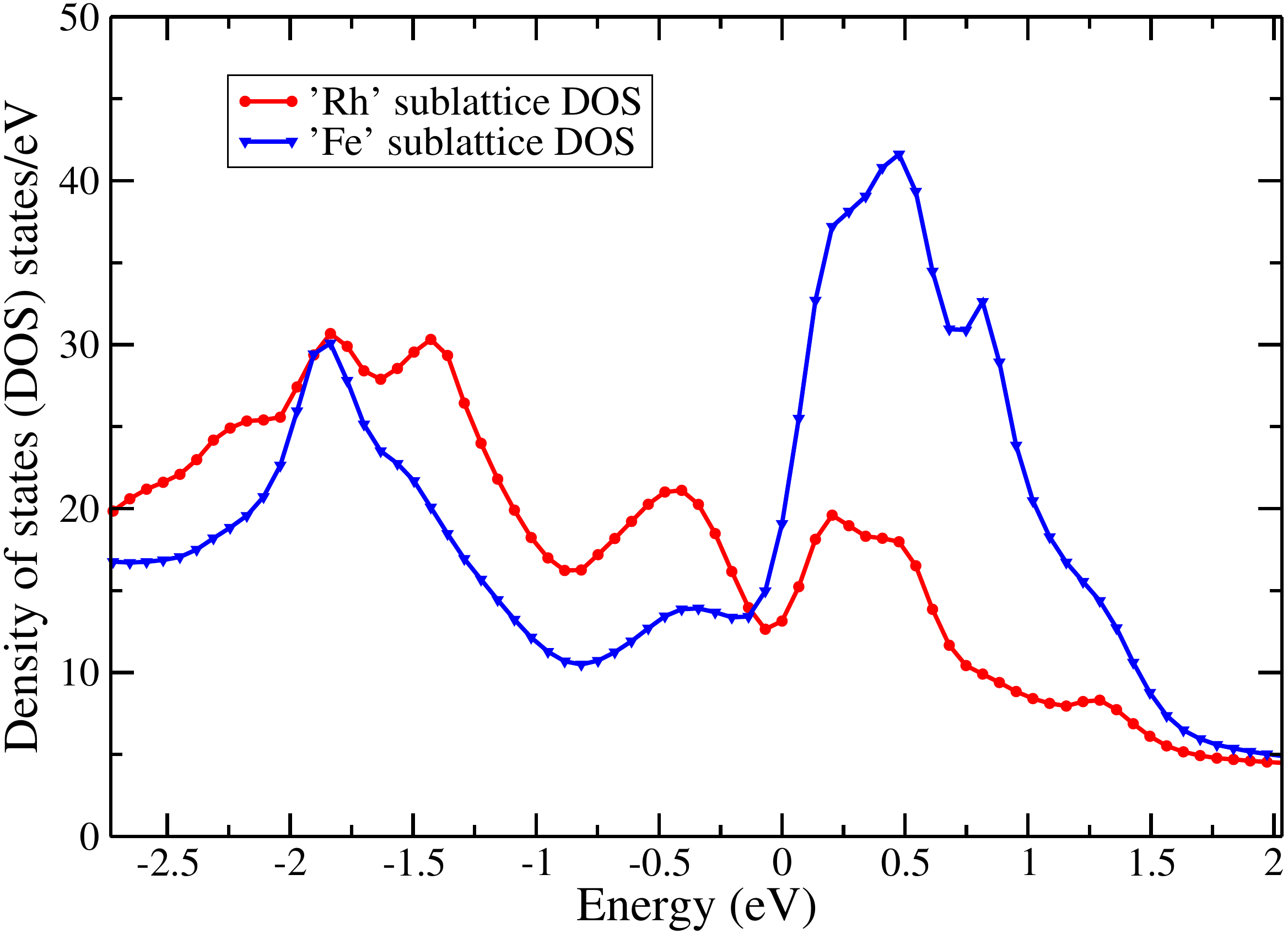}
\includegraphics[width=42mm,height=36mm]{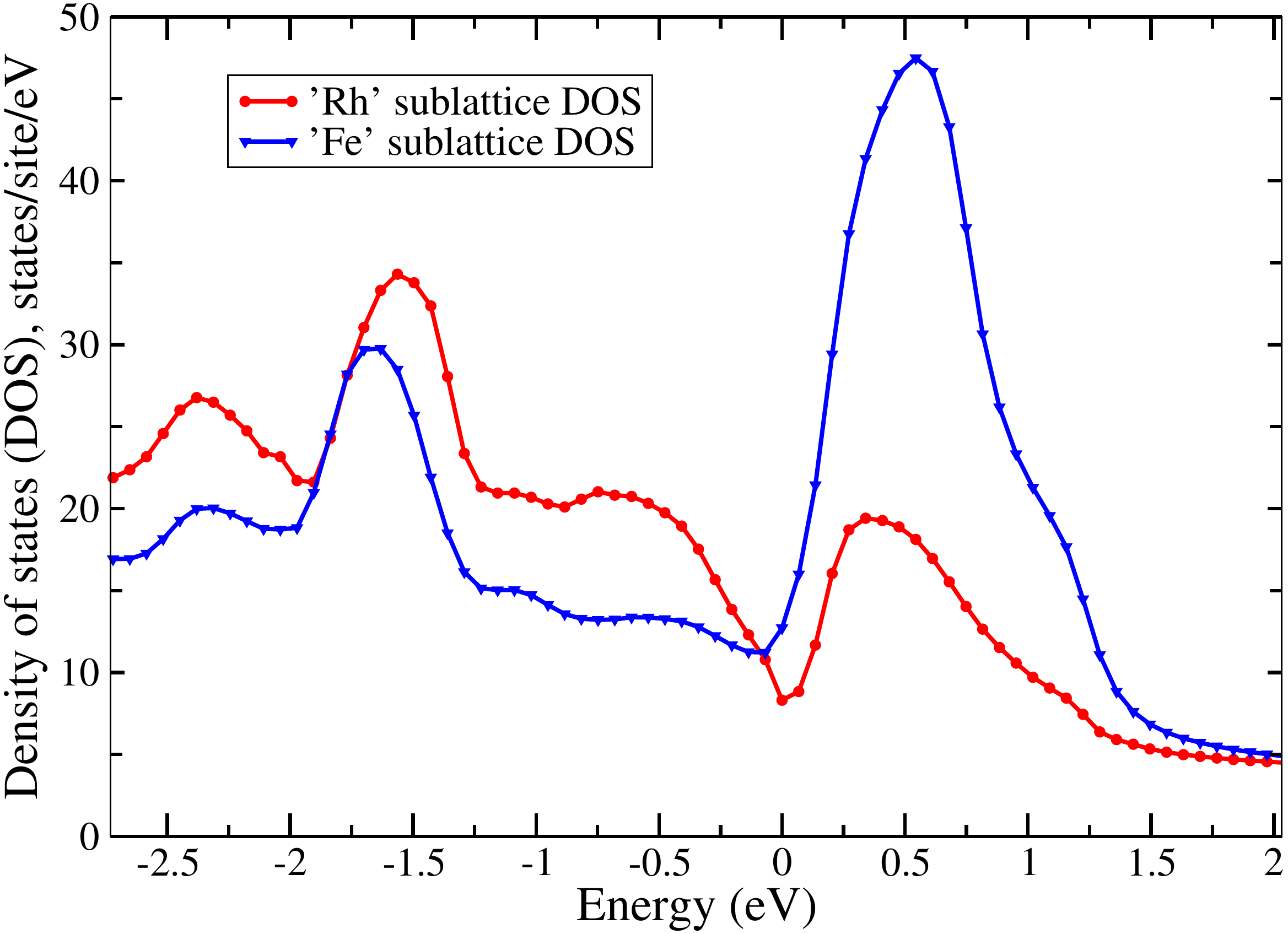}
\caption{(a) Density of states of FeRh just above $T_t$ in the F state, $m_{f} = 0.59$, and (b) just below  $T_t$ in the AF state, $m_{af} = 0.71$.}
\label{Fig1}
\end{center}
\end{figure}

Fig.~1(a) shows the sublattice-resolved scalar-relativistic~\cite{Bordel2012} electronic structure for FeRh above $T_c$ ($m_{f} = 0$, $m_{af} = 0$).
In the paramagnetic state, the Fe atoms still retain a local exchange splitting~\cite{blg} that underpins the stability of the disordered local moments although, once averaged over all the orientations equally weighted, there is no net spin-polarisation in the system.  
In a F state, however, $m_{f} \neq 0$ and the Rh-related states become spin-polarised by the overall long-range alignment of the Fe local moments. 
This lowers the bonding states slightly, as shown in Fig.~2(a) for $T$ just above $T_t$, in comparison with what happens when the Rh states are unpolarised in an AF state ($m_{af} \neq 0$) as shown in Fig.~2(b) for $T$ just below $T_t$.
For the F state there is a DOS peak 1.6 eV below $\EF$ which is slightly lower than its AF counterpart and one set of higher lying states is mostly filled for majority spin electrons as seen in the double peak structure straddling $\EF$ at $\pm$ 0.3eV.
This energy gain is overturned for more entrenched AF order, also evident in Fig. 2(b).
For large $m_{af}$, the AF-coupling between Fe local moments causes a redistribution of states so that a rather pronounced trough develops in the DOS at $\EF$.
The differing effect of F and AF order on this redistribution~\cite{vanDriel1999,Gray2012,Marrows2013} generates the large electronic MCE component and underpins the transition.

Our most important result is the profound effect that tiny compositional changes have on this electronic structure balancing act.
When we model slightly incomplete B2 order by swapping just 2\% of Fe with Rh  ($x = y = 0.02$) we find  $T_t$ to plummet to 208K as found experimentally
~\cite{Tu1969} and $T_c$ to increase to 859K.
For a 4\% swap the F-AF transition vanishes completely ($T_c = 926$K).
Shifting the composition off-stoichiometric has dramatic consequences too.
No F-AF transition is found for the Fe-rich $x=0$, $y=0.04$ Fe$_{52}$Rh$_{48}$ composition ($T_c = 1008$K) whilst for $x=0.04$, $y=0$ (Fe$_{48}$Rh$_{52}$) $T_t$ increases to 549K ($T_c = 700$K).
For the Fe$_{49}$Rh$_{51}$ composition with slightly imperfect B2 order, consistent with reported annealing temperatures, ($x=0.03$, $y=0.01$) we find $T_t = 415K$ ($T_c = 815K$) and $|\Delta S^{max}| = 20.7$ J~K$^{-1}$~Kg$^{-1}$ at 2T, in good agreement with experiment.

The transition's extreme sensitivity to Fe defects on the Rh sublattice is encapsulated by the relatively large positive value of the $g_{ff'}$ coefficient in Eq.~\ref{fit} which describes the ferromagnetic coupling between Fe atoms on the two sub-lattices ($\approx$ 600 meV).
Fig.~1(b) shows the root electronic cause for this.
With the presence of local spin-polarisation on the Fe `defect' sites on the Rh sub-lattice, developing F order deepens the DOS trough at $\EF$ with near depletion of a set of minority spin states. $T=0$K DFT supercell calculations
of Fe anti-site defects~\cite{Kaneta2011} support this interpretation.
This enhances the material's tendency to F order and the F-AF transition is affected accordingly.
Fe anti-site defects reduce the electronic contribution to the MCE, e.g. for $x=y=0.02$, only 20\% of the $|\Delta S| = 22.3$ J~K$^{-1}$~kg$^{-1}$ is electronic. So magnetotransport properties are also acutely composition-sensitive in this material.

The metamagnetism of Fe-Rh varies with impurity doping~\cite{Lewis2013} and this can also be linked to the Fe anti-site effect.
For example isoelectronic dopants such as Pd and Pt have opposite effects on $T_t$ owing the differing propensities for the dopants to displace Fe atoms onto the Rh sub-lattice.
$T = 0$K calculations~\cite{KKR-review} for  the energy difference between F Fe$_{1-2z}$Pt$_{2z}$-Rh$_{1-y}$Fe$_y$ and Fe$_{1-x}$Rh$_{x}$-Rh$_{1-2z}$Pt$_{2z}$) alloys ($0 < z < 10$\%) show that the `big' Pt atoms preferentially displace Rh atoms to the Fe sub-lattice so that they maximise the number of the smaller Fe nearest neighbors~\cite{NiPt}.
This reduces the spin-polarisation energy gain, strengthening AF ordering and increasing $T_t$.
On the other hand similar calculations for the Pd-doped alloys show Fe and Rh atoms are displaced roughly equally so that some Fe atoms find their way onto the Rh sublattice creating Fe antisite defects.
Consequently $T_t$ drops~\cite{Lewis2013}.

At the F-AF transition there is a significant well-studied volume magnetostriction $\lambda$ of $\approx 8 \times 10^{-3}$~\cite{McKinnon1970}.
We find the leading coefficients $e_{f}$ and $e_{af}$ of Eq.~\ref{fit} to change by 16 and -11 meV respectively for a 1\% 
volume increase reflecting the increasing tendency towards F order with expansion. So the free energy, $\FF \approx \FF_0
+ \FF^{'} \lambda + \frac{1}{2} K \lambda^2$. Using
a bulk modulus $K$ estimate of 2 Mbar~\cite{Moruzzi1992},  $\lambda$ at $T_t$ is $4 \times 10^{-3}$ for perfectly ordered Fe-Rh and $6 \times 10^{-3}$ for imperfectly ordered Fe$_{98}$Rh$_{2}$-Rh$_{98}$Fe$_{2}$ alloy, in fair agreement with experiment, and deduce that volume changes are consequences of the magnetic transition but not major drivers of it. 

We have described an ab-initio DFT-based theory for magnetic materials which gives a quantitative account of a material's properties as function of temperature and applied magnetic field.
The role of temperature-dependent spin-polarised electronic structure is paramount.
The unusual ordered FeRh material with its famous metamagnetic transition has all main features described well in comparison with experiment: $\bullet$ the location of the F-AF transition close to room temperature, $\bullet$ the magnitude of the magnetocaloric entropy changes with large electronic contributions
and $\bullet$ crucially the extremely narrow compositional range for the first order transition and its dependence on material preparation route.
In particular we show why very small variations away from complete B2-order and stoichiometry have such a strong effect on the transition in terms of the presence of Fe-antisites, i.e. Fe atoms occupying a very small proportion of the Rh sublattice sites.
In addition to affecting phase coexistence and the broadening of the first order magnetic transition, this compositional hypersensitivity means that, even if Rh were cheaply available, FeRh is unlikely to become a technologically useful adaptive magnetic material.
On the other hand the theory's facility to quantify subtle compositional and electronic effects means that it can aid the design of the next generation of adaptive magnetic materials including magnetic refrigerants based on, for example, MnFePSi or LaFeSi. 
 
Support is acknowledged from the EPSRC (UK) grant EP/J006750/1 (J.B.S. and R.B.), the HGF-YIG 
Programme VH-NG-717 (Functional Nanoscale Structure and Probe Simulation Laboratory) (M. S. D.) and the European Union and the State of Hungary,
co-financed by the European Social Fund, in the framework of TÁMOP-4.2.4.A/1-11/1-2012-0001 ‘National Excellence Program (A. D.).

\end{document}